\documentclass{css2018m}

\usepackage{epsfig}

\newcommand{\PACS}{\MSC}

\title{Can the Dark-Matter Deficit in the High-Redshift Galaxies Explain
       the Persistent Discrepancy in Hubble Constants?}

\author{Yurii V. Dumin$^{1,2}$\\
\vskip 2mm {\small
$^1$P.K.~Sternberg Astronomical Institute
of M.V.~Lomonosov Moscow State University \\
Universitetskii prosp., 13, 119234, Moscow, Russia \\
$^2$Space Research Institute of the Russian Academy of Sciences \\
Profsoyuznaya str.\ 84/32, 117997, Moscow, Russia \\
dumin@yahoo.com, dumin@sai.msu.ru
}}

\abstract{
One of hot topics in the last years is a systematic discrepancy in the
determination of Hubble parameter by various methods. Namely, the values
derived ``directly'' from the distance scale based on Cepheids and
supernovae---and referring to the relatively ``local'' part of the
Universe---are about 10{\%} greater than the ones following from the
analysis of the cosmic microwave background (CMB) radiation,
which refers to the ``global'' scales. The most popular interpretation of
this discord, widely discussed nowadays, is variation of the dark-energy
equation-of-state parameter~$ w $. However, there might be a much simpler
explanation, following from the recent observations of the rotation curves
in the high-redshift galaxies. Namely, it was found that they have much
smaller dark-matter halos than galaxies in the vicinity of us
[Genzel, et al. Nature \textbf{543} (2017), 397]. Since both
the dark and luminous matter possess the same dust-like equation of state
and, therefore, their average cosmological densities evolve by the same
way, our local neighborhood is not quite typical but rather overfilled with
the dark matter. Then, the local value of the Hubble constant should be
greater than the global one. Roughly speaking, a twofold excess of the
dark matter in our local Universe would give just the above-mentioned 10{\%}
increase in the value of Hubble parameter.
}


\begin{document}

\maketitle

Determination of the Hubble parameter~$ H_0 $ is a long-standing problem
in cosmology, lasting for almost a century; and the corresponding values
varied in this period by an order of magnitude,
50 to 500~km\,s$^{-1}$\,Mpc$^{-1}$
(see, for example,~\cite{Freedman_2017,Ryden_2017} and references therein).
Despite of considerable improvements, some discrepancies persist till now.
The most notable of them is that the value of~$ H_0 $ derived from
the distance scale based on Cepheids and supernovae is
$ 73.24{\pm}1.74 $~km\,s$^{-1}$\,Mpc$^{-1}$ and, for some calibration,
can even be as large as
$ 76.18{\pm}2.37 $~km\,s$^{-1}$\,Mpc$^{-1}$~\cite{Riess_2016}.
On the other hand, the analysis based on measurements of the cosmic microwave
background (CMB) by \textit{Planck} satellite under assumption of
the $\Lambda$CDM cosmological model leads to the values
$ H_0 = 66.88{\pm}0.91 $ to $ 67.31{\pm}0.96 $~km\,s$^{-1}$\,Mpc$^{-1}$,
depending on the method of data processing~\cite{Aghanim_2016}.
So, these numbers are about 10{\%} less than in the first case.

The above-mentioned discrepancy between the ``local'' (by Cepheids) and
``global'' (by CMB) measurements of~$ H_0 $ was clearly recognized in
the recent years, and it is commonly attributed now either to
the systematic errors (such as degeneracy between different quantities in
the analysis of CMB) or to the uncertainty in the fitting parameters (e.g.,
the number and masses of neutrinos, etc.)~\cite{Verde_2013,Bernal_2016}.
Yet another popular explanation is a modification of the dark-energy
equation-of-state parameter $ w $ (where $ p = w \rho $)
\cite{Huang_2016,Liu_2016,Di_Valentino_2017,Zhao_2017};
though the resulting values $ w < -1 $ look quite suspicious from
the viewpoint of general physical principles.%
\footnote{For example, the values of~$ w $ somewhat greater than~$ -1 $
(i.e., $ |w| < 1 $) could be easily attributed to the small-scale
irregularities of the scalar field representing the ``dynamic'' dark
energy~\cite{Linde_1984}, but such an effect cannot result in $ w < -1 $.}

However, from our point of view, the spread in values of~$ H_0 $ can have
a much more straightforward explanation, following from the recent
observations of the rotation curves in distant
galaxies~\cite{Genzel_2017,Swinbank_2017}: it was found that the amount of
dark matter is considerably less in the vicinity of galaxies located at large
redshifts, $ z = 0.6{-}2.6 $.
Next, it should be kept in mind that due to the same dust-like equation of
state ($ w \approx 0 $) both for the luminous and dark matter, the ratio of
their densities does not change with cosmological time.
So, we have to conclude that this ratio should be substantially variable
in space and, thereby, the Hubble parameter should be scale-dependent.

\begin{figure}[t]
\begin{center}
\includegraphics[width=10cm]{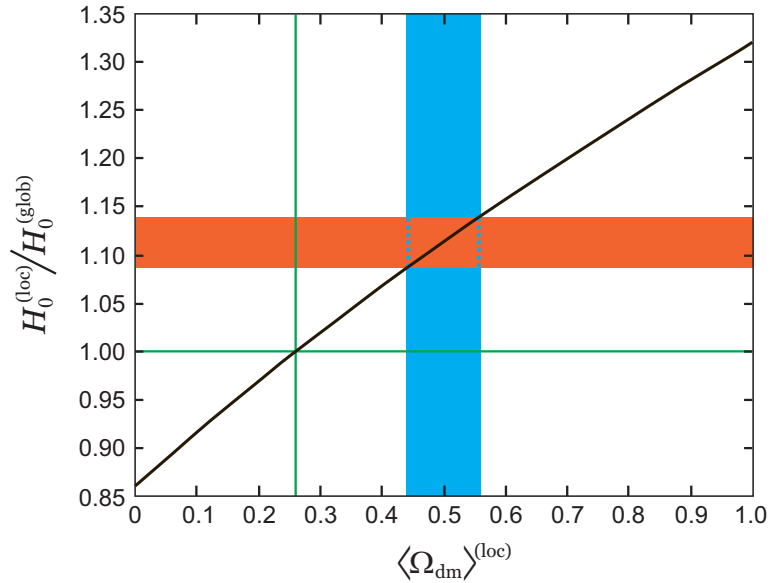}
\caption{\label{fig:1}
Ratio of the Hubble parameters at the local and global scales
$ H_0^{\rm (loc)} / H_0^{\rm (glob)} $ as function of the local
dark-matter density normalized to the global critical density
$ {\langle} {\Omega}_{\rm dm} {\rangle}^{\rm (loc)} $ (diagonal black curve).
Horizontal red and vertical blue strips show the range of observable values.
Horizontal and vertical green lines correspond to the trivial case when
$ {\langle} {\Omega}_{\rm dm} {\rangle}^{\rm (loc)} =
{\langle} {\Omega}_{\rm dm} {\rangle}^{\rm (glob)} $.}
\end{center}
\end{figure}

Really, according to the standard Friedmann equation~\cite{Olive_2016}:
\begin{equation}
H_0 = {\Bigg[} \frac{8 \pi G}{3} {\Bigg]}^{1/2}
      {\Big[} {\rho}_{\rm de} + \langle {\rho}_{\rm dm} \rangle +
      \langle {\rho}_{\rm lm} \rangle {\Big]}^{1/2} ,
\label{eq:Hubble_param}
\end{equation}
where $ {\rho}_{\rm de} $~is density of the dark energy, which is assumed
to be perfectly uniform in space (i.e., described by the $ \Lambda $-term),
$ {\rho}_{\rm dm} $ and $ {\rho}_{\rm lm} $~are densities of the dark
and luminous (baryonic) matter,
and the angular brackets denote averaging over the given spatial scale.
Then, ratio of the Hubble parameters at the local and global scales
should be:%
\footnote{For simplicity, we ignore here the curvature term that might appear
at the local scales due to the non-uniform dark matter distribution.}
\begin{equation}
\frac{H_0^{\rm (loc)}}{H_0^{\rm (glob)}} =
  {\Bigg[} \frac{{\Omega}_{\rm de} +
    {\langle} {\Omega}_{\rm dm} {\rangle}^{\rm (loc)} +
      {\langle} {\Omega}_{\rm lm} {\rangle}}%
  {{\Omega}_{\rm de} + {\langle} {\Omega}_{\rm dm} {\rangle}^{\rm (glob)} +
    {\langle} {\Omega}_{\rm lm} {\rangle}} {\Bigg]}^{1/2} ,
\label{eq:ratio_Hubble_param}
\end{equation}
where $ {\Omega}_i = {\rho}_i / {\rho}_{\rm c} $~are the corresponding
densities normalized to the critical density at the global scale; and
we assume that the luminous matter distribution is sufficiently uniform.

This ratio of the Hubble parameters is plotted in Fig.~\ref{fig:1} as
function of the local dark-matter density at the standard cosmological
parameters:
$ {\Omega}_{\rm de} = 0.69 $,
$ {\langle} {\Omega}_{\rm dm} {\rangle}^{\rm (glob)} = 0.26 $, and
$ {\langle} {\Omega}_{\rm lm} {\rangle} = 0.05 $.
The range of observed values of $ H_0^{\rm (loc)} / H_0^{\rm (glob)} $
is shown by the horizontal red strip.
Then, the corresponding normalized densities of the dark matter in
our local cosmological neighborhood should be in the range
$ {\langle} {\Omega}_{\rm dm} {\rangle}^{\rm (loc)} =  0.44-0.56 $
(vertical blue strip), i.e. about two times greater than globally.

In fact, Genzel, et al.~\cite{Genzel_2017} already emphasized that
at the global scales the dark matter should play a smaller part than
in the local Universe.
So, from our point of view, the systematic discrepancy between the
``local'' and ``global'' values of the Hubble parameter is just
a direct consequence of the above-mentioned fact.

Finally, let us mention that a number of papers published in the recent
years made just the opposite statement as compared to~\cite{Genzel_2017}:
namely, that there is a considerable deficit of luminous and dark matter
in our local cosmological neighborhood.
For example, Makarov~\& Karachentsev~\cite{Makarov_2011} and
Karachentsev~\cite{Karachentsev_2012} found
$ {\Omega}_{\rm dm} \! + {\Omega}_{\rm lm} = 0.08{\pm}0.02 $
in the sphere of radius $ z~{\sim}~0.01 $ around us, which is over
three times smaller than the standard value in the $\Lambda$CDM model.
Unfortunately, their analysis involved a lot of model assumptions.
On the other hand, the work by Genzel, et al.~\cite{Genzel_2017}, which
is based solely on the galaxy rotation curves, seems to be much less
model-dependent; and the corresponding results on the deficit of dark
matter in the high-redshift (rather than local) galaxies look more
reliable.

\section*{Acknowledgements}

I am grateful to Yu.V.~Baryshev, S.M.~Kopeikin, M.~K{\v r}{\'i}{\v z}ek,
A.~Maeder, and M.~Nowakowski for valuable discussions of the problem of
small-scale Hubble expansion, as well as to the referee for a few
important remarks.


\begin{thebibliography}{99}

\bibitem{Freedman_2017}
Freedman,~W.L.:
Cosmology at a crossroads.
Nature Astron. \textbf{1} (2017), 0121.

\bibitem{Ryden_2017}
Ryden,~B.:
A constant conflict.
Nature Phys. \textbf{13} (2017), 314.

\bibitem{Riess_2016}
Riess,~A.G., Macri,~L.M., Hoffmann,~S.L., et al.:
A 2.4\% determination of the local value of the Hubble constant.
Astrophys.\ J. \textbf{826} (2016), 56.

\bibitem{Aghanim_2016}
Aghanim,~N., Ashdown,~M., Aumont,~J., et al. (Planck Collaboration):
Planck intermediate results: XLVI. Reduction of large-scale systematic effects
in HFI polarization maps and estimation of the reionization optical depth.
Astron.\ Astrophys. \textbf{596} (2016), A107.

\bibitem{Verde_2013}
Verde,~L., Protopapas,~P., and Jimenez,~R.:
Planck and the local Universe: Quantifying the tension.
Phys.\ Dark Univ. \textbf{2} (2013), 166.

\bibitem{Bernal_2016}
Bernal,~J.L., Verde,~L., and Riess,~A.G.:
The trouble with $ H_0 $.
J.\ Cosmol.\ Astropart.\ Phys. \textbf{10} (2016), 019.

\bibitem{Huang_2016}
Huang,~Q.-G. and Wang,~K.:
How the dark energy can reconcile Planck with local determination of
the Hubble constant.
Eur.\ Phys.\ J.\ C \textbf{76} (2016), 506.

\bibitem{Liu_2016}
Liu,~Z.-E., Yu,~H.-R., Zhang,~T.-J., and Tang,~Y.-K.:
Direct reconstruction of dynamical dark energy from observational
Hubble parameter data.
Phys.\ Dark Univ. \textbf{14} (2016), 21.

\bibitem{Di_Valentino_2017}
Di~Valentino,~E.:
Crack in the cosmological paradigm.
Nature Astron. \textbf{1} (2017), 569.

\bibitem{Zhao_2017}
Zhao,~G.-B., Raveri,~M., Pogosian,~L., et al.:
Dynamical dark energy in light of the latest observations.
Nature Astron. \textbf{1} (2017), 627.

\bibitem{Linde_1984}
Linde,~A.D.:
The inflationary Universe.
Rep.\ Prog.\ Phys. \textbf{47} (1984), 925.

\bibitem{Genzel_2017}
Genzel,~R., F{\"o}rster Schreiber,~N.M., {\"U}bler,~H., et al.:
Strongly baryon-dominated disk galaxies at the peak of galaxy formation
ten billion years ago.
Nature \textbf{543} (2017), 397.

\bibitem{Swinbank_2017}
Swinbank,~M.:
Distant galaxies lack dark matter.
Nature \textbf{543} (2017), 318.

\bibitem{Olive_2016}
Olive,~K.A. and Peacock,~J.A.:
Big-bang cosmology.
In: Particle Data Group:
\emph{Review of Particle Physics.}
Chin.\ Phys.\ C \textbf{40} (2016), 100001, p.~355.

\bibitem{Makarov_2011}
Makarov,~D. and Karachentsev,~I.:
Galaxy groups and clouds in the local ($ z~{\sim}~0.01 $) Universe.
Mon.\ Not.\ R.\ Astron.\ Soc. \textbf{412} (2011), 2498.

\bibitem{Karachentsev_2012}
Karachentsev,~I.D.:
Missing dark matter in the local Universe.
Astrophys.\ Bull. \textbf{67} (2012), 123.

\end{thebibliography}
\end{document}